\documentclass[11pt]{article}
\usepackage{lscape,epsfig,amsmath,amssymb,latexsym,
	psfig,psfrag,graphicx}
\usepackage{a4wide}
\usepackage{colordvi,gnuplotcolors}

\newcommand{\gev}{\text{GeV}}

\newcommand{\pb}{\text{pb}}

\newcommand{\mhpm}{ m_{H^{\pm}} }

\newcommand{\MSF}{ M_{\text{Sf.}}}

\newcommand{\tb}{\tan\beta}

\newcommand{\TE}[1]{\cdot 10^{#1}}
\newcommand{\E}[1]{10^{#1}}

\newcommand{\MIN}{{\text{min}}}
\newcommand{\MAX}{{\text{max}}}

\begin{document}

\title{
\noindent \hfill {\small hep-ph/0407340}\\
\vspace*{-.3cm}
\hfill {\small PITHA 04/12}\\
\vspace*{.2cm}
Adaptive scanning -- a proposal how to scan theoretical 
predictions over a multi-dimensional parameter space
efficiently
}
\author{{\bf Oliver Brein}\footnote{E-Mail: brein@physik.rwth-aachen.de}\\[.2cm]
{\em Institut f\"ur Theoretische Physik E, RWTH Aachen,
    D-52056 Aachen, Germany}
}

\date{\vspace*{-1cm}}

\maketitle

\begin{abstract}
A method is presented
to exploit adaptive integration algorithms using 
importance sampling, like VEGAS,
for the task of scanning theoretical predictions depending on 
a multi-dimensional parameter space.
Usually, a parameter scan is performed with
emphasis 
on certain features of a theoretical
prediction.
Adaptive integration algorithms are well-suited to perform
this task very efficiently.
Predictions 
which depend on parameter spaces with many dimensions
call for such an adaptive scanning algorithm.
\end{abstract}

\section{Introduction}

Most observations in elementary particle physics so far are well described 
by the Standard Model of particle physics. However, it is widely believed
that the Standard Model is only an effective low-energy limit embedded in 
a larger theory which is not known at present.
Extensions of the Standard Model of particle physics usually are 
equipped with a lot of additional parameters compared to the 
Standard Model which describe new physics
at energy scales not probed so far by experiment.
The predictions of promising extensions of the Standard Model 
for physical observables should be calculated and studied in detail,
either to find hints for new physics or to exclude
alternative models or, at least, to constrain their parameters.

A way to study such predictions is performing parameter scans.
But what exactly is a parameter scan? 
A parameter scan may be defined as the process of calculating numerical values
for a theoretical prediction for a sufficient amount of points in parameter
space, such that, sufficient smoothness of the prediction's dependence 
on the parameters assumed, a clear understanding of the range of values
for the prediction arises.
Usually, in this process all other known restrictions on the parameters
are taken into account. Thus, only meaningful, i.e.
not otherwise excluded, points in parameter space are sampled.
In high energy physics
the most common goal of a parameter scan over a prediction of an alternative
model 
is to restrict the parameter space of the model by comparison 
of the prediction
with experimental results.
Specifically, one wants to find the regions in parameter space where a
theoretical prediction
for an observable either matches the measured value within error bars,
or, if nothing has been observed,
respects
exclusion limits on possible observables drawn from
experiments.
This is a typical situation.
In practice, of course, 
things are 
sometimes more involved.

Common techniques for parameter scans are 
sampling
a rigid grid of points
in parameter space or 
choosing
points at random with a uniform distribution.
For the latter sometimes special non-uniform probability functions are used
for plotting reasons, e.g. to have the sample points uniformly distributed on 
a logarithmic scale.
But, in the same way as a rigid grid of samples or the naive Monte Carlo approach
are not the best options for evaluating numerically multi-dimensional
integrals efficiently these parameter scan techniques are not optimal for
multi-dimensional parameter spaces.
The amount of calculation time needed for a parameter scan may 
easily get out of hand for parameter spaces with many dimensions,
for instance
the minimal supersymmetric standard model (MSSM) with general parameters
---
especially, if time-consuming calculations are necessary
already for each sampled parameter point, 
as for instance
in many predictions for hadron colliders.

The basic observation made 
in this note
is that finding regions of
interest for a function in a multi-dimensional parameter space and 
finding a good approximation to the integral of a multi-dimensional function
can be attacked by the same method 
in a very efficient way.
That means, essentially, existing algorithms
for adaptive integration by importance sampling
(implemented e.g. in VEGAS \cite{vegas})
can also be used to perform an 
what will here be called
``adaptive parameter scan''.

In section \ref{algorithm} a formulation of our proposal is given, furnished 
in section \ref{examples} with a concrete example from the study
of an MSSM cross section prediction, and summarized in section \ref{summary}.

\section{Algorithm}\label{algorithm}

By an adaptive parameter scan we mean the task to scan a function $f$,
depending on a set of $N$ parameters $(p_1,\ldots,p_N)$, 
over a hypercube\footnote{As is well known from adaptive Monte Carlo
integration algorithms, any other, more complicated, parameter region can always
be embedded in a hypercube with little loss of efficiency.}
$\{(p_1,\ldots,p_N) \,|\, p_i^\MIN\leq p_i \leq p_i^\MAX, i=1,\ldots,N\}$
in order to find regions
of parameter space where some ``importance function'' $G(f,p_1,\ldots,p_N)$ becomes large.
To that end an approximation to the integral
\begin{align}
I& =
\int_{p_1^\MIN}^{p_1^\MAX}\! dp_1 \cdots \int_{p_N^\MIN}^{p_N^\MAX} \! dp_N \,
G(f(p_1,\ldots,p_N),p_1,\ldots,p_N)
\end{align}
is evaluated by an adaptive integration algorithm using importance sampling
(e.g. VEGAS) and
each point in parameter space is stored during the process.
\footnote{It should be mentioned that
a completely different form of a parameter scan appears 
in \cite{ellwanger}
which also makes use of an integration over parameter space.
}
The idea is to choose $G$ in such a way that the (for some specific reasons)
preferred
values of $f$ (and possibly $p_i$) lead to large values of $G$ and that therefore 
the parameter regions where $G(f,p_i)$ 
is large (i.e. where $f$ and possibly the $p_i$ take on preferred values
according to the specific reasons which lead to the choice of $G$)
are more thoroughly sampled by an adaptive integration algorithm based on
importance sampling.
Examples are:
\begin{itemize}
\item $G(f)=f$ to emphasize regions with large values of $|f|$,
\item or $G(f)=1/f$ to emphasize regions with small values of $|f|$,
\item or $G(f)=\Theta(-f) f$ to emphasize regions with large $|f|$ but with $f$
negative.
\end{itemize}
The end result is a set of points in parameter space which may allow
to give meaningful answers to questions like 
\begin{itemize}
\item ''What are the regions of large cross section ?'', 
\item ''Where are the regions in parameter space where 
the observable is small enough to be consistent with present data ?'', 
\item ''In what regions of parameter space
are the radiative corrections to some observable negative ?''.
\end{itemize}
If there are complicated parameter restrictions to be taken into account
the method proves to be very powerful. By setting the importance function
zero in disallowed regions and non-zero otherwise, the algorithm 
can adapt to the allowed region of parameter space. 
Thus, even if 
no other feature of the theoretical prediction 
is emphasized (i.e. $G(f)=1$ in the allowed region),
the 
algorithm focuses iteratively on the allowed region, which is not 
the case in the naive Monte Carlo approach.

For the integration of a function by an adaptive Monte Carlo algorithm
the approximate value and error of the integral estimate is 
most important, while the sampled points are essentially of no interest.
For adaptive scanning it's just the opposite. Most interesting are the 
sampled points, while the approximate value of the estimate and its error 
are quantities to control the thoroughness of the scan.

\section{Example}\label{examples}

In order to demonstrate the usefulness of the proposed method we apply 
it to a concrete physics example based on a previous calculation \cite{eewh}.
Parts of these results have been already presented elsewhere
\cite{chargedHiggs}. We want to study the MSSM prediction for the cross section 
$\sigma$ of the process
$e^+ e^- \to H^\pm W^\mp$, where $H^\pm$ denotes a charged Higgs boson
and $W^\mp$ a charged electroweak vector boson. 
For our example we 
choose a center--of--mass energy $\sqrt{s} =500\,\gev$ 
for the colliding $e^+ e^-$ pair, which is typical for a future  high energy 
linear collider, and for the mass of the charged Higgs boson $\mhpm = 250\,\gev$.
In this situation the production of charged Higgs pairs would just be 
kinematically forbidden and the process under study would be one of
the relevant production channels for the charged Higgs search at a
linear collider of this energy. 
The cross section of this loop-induced process is rather small. 
Therefore, it is interesting to scan the MSSM parameter space
with an emphasis on regions with comparatively large cross section.

In the following we compare the results of 
a parameter scan performed in three different ways.
For simplicity we scan over a 6-dimensional subset of the MSSM parameter space
in the ranges
\begin{align*}
1 & \leq \tb \leq 50 \:,& 
-2000\,\gev & \leq \mu, A_t, A_b \leq 2000\,\gev \:,& 
50\,\gev & \leq \MSF, M_2 \leq 2000\,\gev \:,
\end{align*}
where $\tb$ is the ratio of the two Higgs vacuum expectation values in the
MSSM,
$\mu$ is the supersymmetric Higgs mass term,
$A_t$ and $A_b$ are soft-breaking trilinear couplings,
$\MSF$ is a common sfermion mass scale 
and $M_2$ a common gaugino mass scale.
The other two gaugino mass parameters $M_1$ and $M_3$ are fixed 
by GUT relations. 
A more detailed description of the parameters and
relations can be found in \cite{schappi-hahn}, which has been used 
for the calculation of the cross section together with 
the computer programs FeynArts and FormCalc \cite{FAFC}.
In order to restrict our scan to meaningful combinations of parameters 
we discard all points in parameter space where 
experimental bounds on masses of so far undiscovered particles, like
Higgs bosons, sfermions and gauginos, are violated, and where 
corrections to the electroweak rho-parameter by
superpartner loop-contributions strongly disagree 
with the measured value. 
We choose the bounds listed in \cite{schappi-hahn}, which might
seem too conservative for a real physics analysis 
but are sufficient for our demonstration purpose.
For all adaptive scans discussed in the following the importance function
$G$ was set to zero for discarded points, thus influencing the adaptive
scan.

\begin{figure}[t]
{\setlength{\unitlength}{1cm}
\begin{picture}(14,4.2)(0,0)
  \psfrag{naive}[l][l]{\LARGE naive}
  \psfrag{ADGISSIGMA}[l][l]{\LARGE adaptive, $G=\sigma$}
  \psfrag{ADGISTHETASIGMA}[l][l]{\LARGE adaptive, $G=\theta(\sigma-\sigma_0) \sigma$}
  \psfrag{M_sfermion}[c][b]{\LARGE $\MSF [\gev]$} 
  \psfrag{TANB}[c][b]{\LARGE $\tb$}
\put(.5,0){\resizebox*{.348\width}{.348\height}{
\includegraphics*{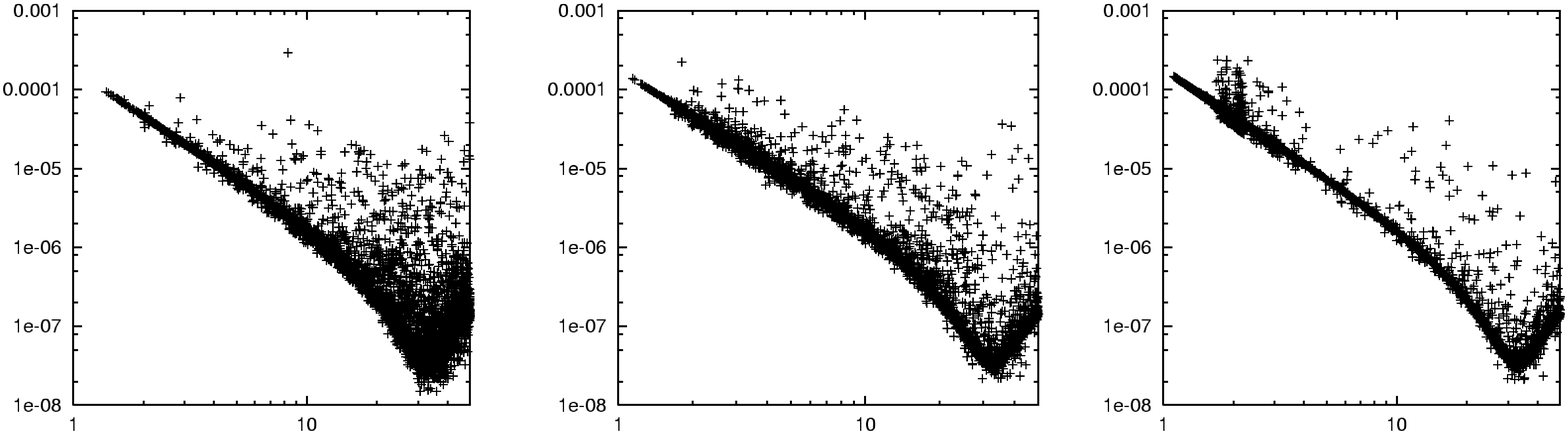}}}
\put(4.0,3.65){\tiny naive}
\put(8.,3.65){\tiny adaptive, $G=\sigma$}
\put(11.8,3.65){\tiny adaptive,{\tiny $G=\theta(\sigma-\sigma_0)\,\sigma$}}
\put(.5,3.52){$\scriptstyle\sigma\; [{\text{\tiny pb}}]$}
\put(4.3,.1){$\scriptstyle\tb$}
\put(0,2){(a)}
\end{picture}
}

{\setlength{\unitlength}{1cm}
\begin{picture}(14,4.2)(0,0)
  \psfrag{naive}[l][l]{\LARGE naive}
  \psfrag{ADGISSIGMA}[l][l]{\LARGE adaptive, $G=\sigma$}
  \psfrag{ADGISTHETASIGMA}[l][l]{\LARGE adaptive, $G=\theta(\sigma-\sigma_0) \sigma$}
  \psfrag{M_sfermion}[c][b]{\LARGE $\MSF [\gev]$} 
  \psfrag{TANB}[c][b]{\LARGE $\tb$}
\put(.5,0){\resizebox*{.348\width}{.348\height}{\includegraphics*{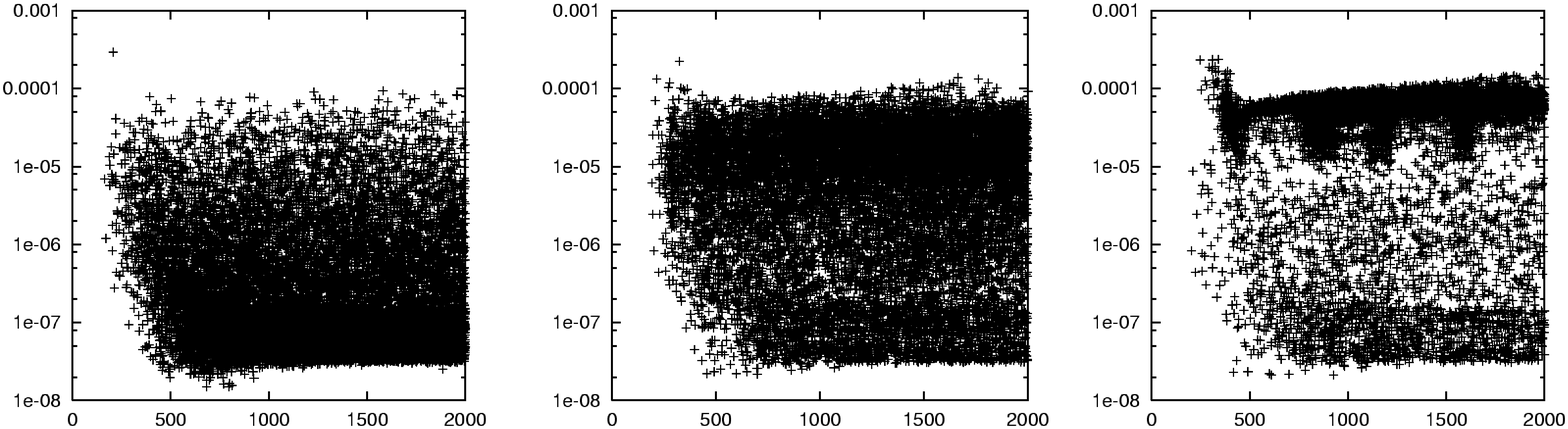}}}
\put(4.0,3.65){\tiny naive}
\put(8.,3.65){\tiny adaptive, $G=\sigma$}
\put(11.8,3.65){\tiny adaptive,{\tiny $G=\theta(\sigma-\sigma_0)\,\sigma$}}
\put(.5,3.52){$\scriptstyle\sigma\; [\text{\tiny pb}]$}
\put(3.5,-.1){$\scriptstyle\MSF\; [\text{\tiny GeV}]$}
\put(0,2){(b)}
\end{picture}
}

{\setlength{\unitlength}{1cm}
\begin{picture}(14,4.2)(0,0)
  \psfrag{naive}[l][l]{\LARGE naive}
  \psfrag{ADGISSIGMA}[l][l]{\LARGE adaptive, $G=\sigma$}
  \psfrag{ADGISTHETASIGMA}[l][l]{\LARGE adaptive, $G=\theta(\sigma-\sigma_0) \sigma$}
  \psfrag{M_sfermion}[c][b]{\LARGE $\MSF [\gev]$} 
  \psfrag{TANB}[c][b]{\LARGE $\tb$}
\put(.5,0){\resizebox*{.348\width}{.348\height}{\includegraphics*{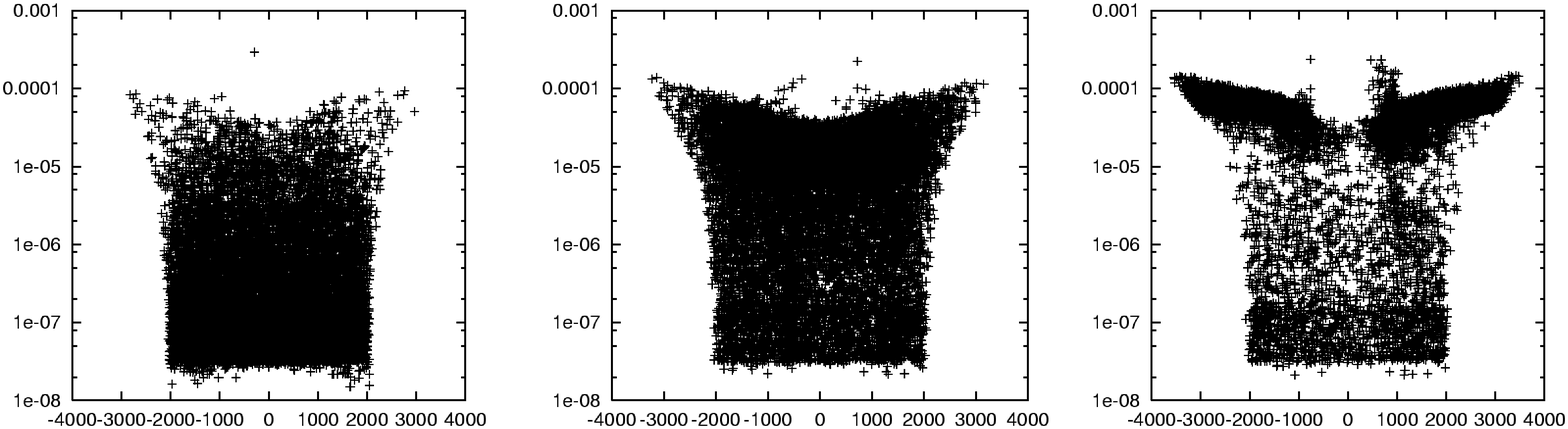}}}
\put(4.0,3.65){\tiny naive}
\put(8.,3.65){\tiny adaptive, $G=\sigma$}
\put(11.8,3.65){\tiny adaptive,{\tiny $G=\theta(\sigma-\sigma_0)\,\sigma$}}
\put(.5,3.52){$\scriptstyle\sigma\; [\text{\tiny pb}]$}
\put(3.5,-.1){$\scriptstyle X_t\; [\text{\tiny GeV}]$}
\put(0,2){(c)}
\end{picture}
}
\caption{\label{1dplots-log}
Cross section values in picobarn obtained by
the three ways of scanning the parameter space 
plotted on a logarithmic scale versus one scan parameter, 
(a) $\tb$, (b) $\MSF$ and (c) $X_t = A_t - \mu/\tb$.}
\end{figure}

\begin{figure}[ht]
{\setlength{\unitlength}{1cm}
\begin{picture}(14,4.2)(0,0)
  \psfrag{naive}[l][l]{\LARGE naive}
  \psfrag{ADGISSIGMA}[l][l]{\LARGE adaptive, $G=\sigma$}
  \psfrag{ADGISTHETASIGMA}[l][l]{\LARGE adaptive, $G=\theta(\sigma-\sigma_0) \sigma$}
  \psfrag{M_sfermion}[c][b]{\LARGE $\MSF [\gev]$} 
  \psfrag{TANB}[c][b]{\LARGE $\tb$}
\put(.5,0){\resizebox*{.348\width}{.348\height}{\includegraphics*{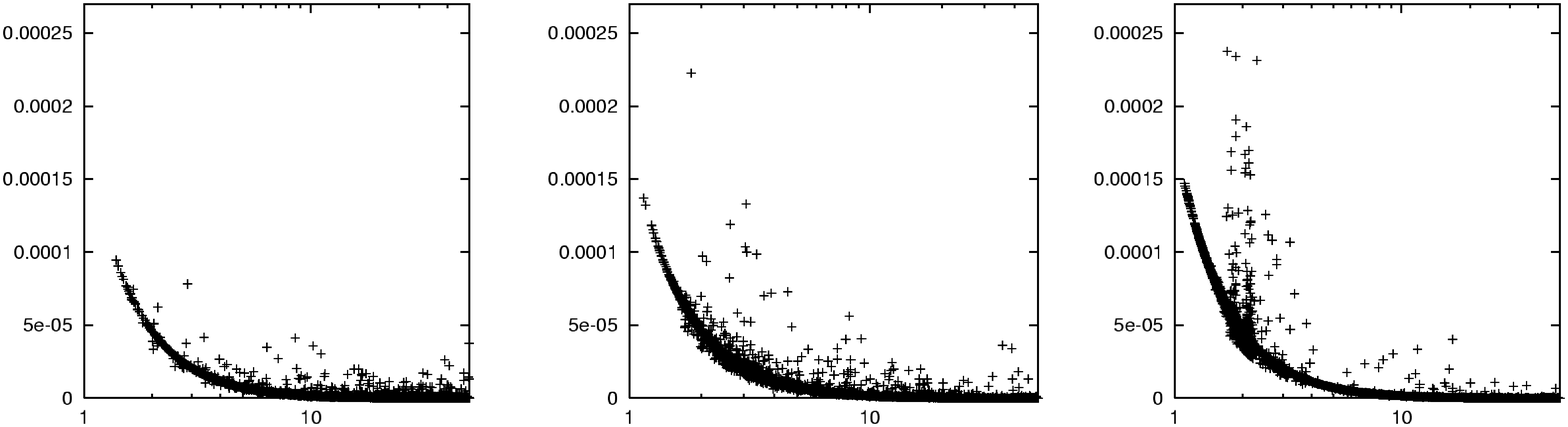}}}
\put(4.0,3.65){\tiny naive}
\put(8.,3.65){\tiny adaptive, $G=\sigma$}
\put(11.8,3.65){\tiny adaptive,{\tiny $G=\theta(\sigma-\sigma_0)\,\sigma$}}

\put(.5,3.3){$\scriptstyle\sigma\; [{\text{\tiny pb}}]$}
\put(4.3,.1){$\scriptstyle\tb$}
\put(0,2){(a)}
\end{picture}
}

{\setlength{\unitlength}{1cm}
\begin{picture}(14,4.2)(0,0)
  \psfrag{naive}[l][l]{\LARGE naive}
  \psfrag{ADGISSIGMA}[l][l]{\LARGE adaptive, $G=\sigma$}
  \psfrag{ADGISTHETASIGMA}[l][l]{\LARGE adaptive, $G=\theta(\sigma-\sigma_0) \sigma$}
  \psfrag{M_sfermion}[c][b]{\LARGE $\MSF [\gev]$} 
  \psfrag{TANB}[c][b]{\LARGE $\tb$}
\put(.5,0){\resizebox*{.348\width}{.348\height}{\includegraphics*{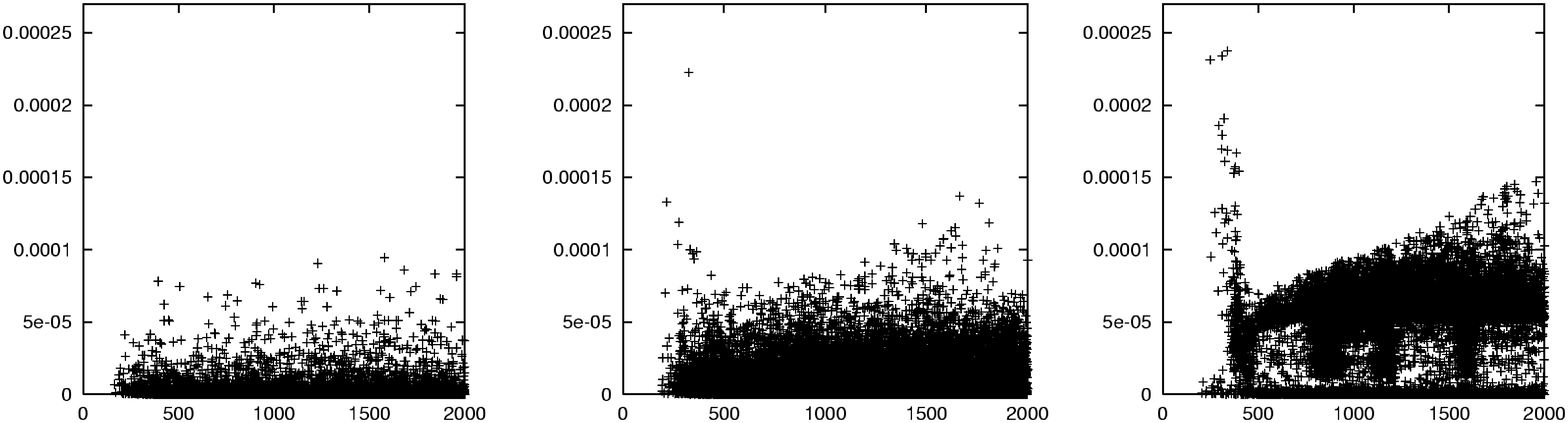}}}
\put(4.0,3.65){\tiny naive}
\put(8.,3.65){\tiny adaptive, $G=\sigma$}
\put(11.8,3.65){\tiny adaptive,{\tiny $G=\theta(\sigma-\sigma_0)\,\sigma$}}
\put(.5,3.3){$\scriptstyle\sigma\; [\text{\tiny pb}]$}
\put(3.5,-.1){$\scriptstyle\MSF\; [\text{\tiny GeV}]$}
\put(0,2){(b)}
\end{picture}
}

{\setlength{\unitlength}{1cm}
\begin{picture}(14,4.2)(0,0)
  \psfrag{naive}[l][l]{\LARGE naive}
  \psfrag{ADGISSIGMA}[l][l]{\LARGE adaptive, $G=\sigma$}
  \psfrag{ADGISTHETASIGMA}[l][l]{\LARGE adaptive, $G=\theta(\sigma-\sigma_0) \sigma$}
  \psfrag{M_sfermion}[c][b]{\LARGE $\MSF [\gev]$} 
  \psfrag{TANB}[c][b]{\LARGE $\tb$}
\put(.5,0){\resizebox*{.348\width}{.348\height}{\includegraphics*{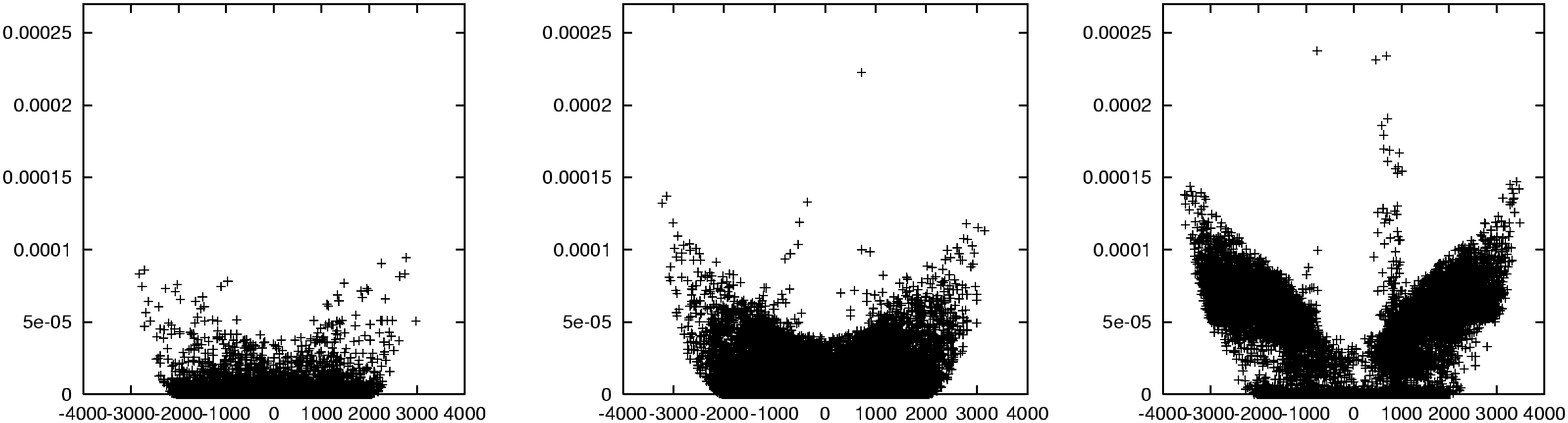}}}
\put(4.0,3.65){\tiny naive}
\put(8.,3.65){\tiny adaptive, $G=\sigma$}
\put(11.8,3.65){\tiny adaptive,{\tiny $G=\theta(\sigma-\sigma_0)\,\sigma$}}
\put(.5,3.3){$\scriptstyle\sigma\; [\text{\tiny pb}]$}
\put(3.5,-.1){$\scriptstyle X_t\; [\text{\tiny GeV}]$}
\put(0,2){(c)}
\end{picture}
}
\caption{\label{1dplots-lin}
Cross section values in picobarn obtained by
the three ways of scanning the parameter space 
plotted on a linear scale versus one scan parameter, 
(a) $\tb$, (b) $\MSF$ and (c) $X_t = A_t - \mu/\tb$.}
\end{figure}

We perform the parameter scan using \\[1mm]
(A) the naive Monte Carlo approach,
i.e. just randomly sampling points in the given hypercube,\\[1mm]
(B) the adaptive Monte Carlo approach using VEGAS with 11 iterations
and 2000 sample points per iteration with the ``regular'' importance function
$G(\sigma) = \sigma$, i.e. integrating $\sigma$ over the hypercube
and recording the sampled points,\\[1mm]
(C) the adaptive Monte Carlo approach using VEGAS with 12 iterations
and 2000 sample points per iteration
with the importance function 
$G(\sigma)= \theta(\sigma - \sigma_0)\cdot\sigma$, where 
$\sigma_0 = 5\TE{-4}\pb$ and $\theta(x)=0$
for $x<0$ and $\theta(x)=1$ for $x>0$.\\[1mm]
The last approach already uses some knowledge about the range of values 
$\sigma$ 
can take on and is a rather radical way to focus on
large values of $\sigma$
by disfavoring all regions with $\sigma < 5\TE{-4}\pb$ by setting
the importance function to zero there.
The VEGAS settings of approaches (B) and (C) are such that the number 
of valid points is approximately equal (14874 and 15466)
and (A) has the same number of points as (C).
Thus, the performance of the three methods can be judged easily
by comparing Figures \ref{1dplots-log} and \ref{1dplots-lin}.
Note that we don't consider a rigid grid of points.
While this might give good results for parameter spaces of low
dimensionality, calculation time will blow up like $n^N$, where 
$n$ is the desired number of sample points per scan parameter and $N$
the number of scan parameters.
Already for 10 points per parameter this would require in our 
example $\E{6}$ points to calculate.

In the following the results 
for the three ways of scanning the parameter space are
discussed.
Figures \ref{1dplots-log} and \ref{1dplots-lin} show 
the cross section 
plotted versus one of the scan parameters
on a logarithmic and linear scale, respectively.
For brevity we don't show plots for all scan parameters.
As interesting examples we plot the cross section versus 
$\tb$ (Figures \ref{1dplots-log}(a) and \ref{1dplots-lin}(a)), 
$\MSF$ (Figures \ref{1dplots-log}(b) and \ref{1dplots-log}(b)) and
versus
$X_t = A_t - \mu/\tb$ (Figures \ref{1dplots-log}(c) and \ref{1dplots-log}(c)). 
The latter combination of 
parameters controls the mixing between interaction- and mass-eigenstates 
in the scalar top sector of the MSSM and is well known to control
the size of many quantum effects in the MSSM.
Figures \ref{1dplots-log}(a) and \ref{1dplots-lin}(a) show that
there is a strong dependence on $\tb$, such that even in the scan results
a slope can be guessed, which most of the cross section values follow closely.
Comparing the different ways of scanning there is clearly a depletion
of points with lower values of $\sigma$ for the adaptive approaches
compared to the naive approach, whereas the enhancement of sampled points
with larger $\sigma$--values are better seen in Figures
\ref{1dplots-log} and \ref{1dplots-lin} (b) and (c).

In Figures \ref{1dplots-log}(b) and \ref{1dplots-lin}(b) the adaptive scan
using the $\theta$--function to disfavor lower values of $\sigma$
(approach (C))
already defines a clear upper boundary of possible $\sigma$-values
over a wide range in $\MSF$. 
Only for $\MSF$ below $\approx 300\,\gev$ more statistics is needed 
to get a clear boundary.

Figures \ref{1dplots-log}(c) and \ref{1dplots-lin}(c) show that the adaptive scans,
(B) and (C),
resolve the areas of large $X_t$ that lead to large cross section, which
is not the case for the naive approach.
The adaptive scans also reveal two areas of large cross section around 
$X_t \approx \pm 1000\,\gev$ which would need more statistics to be 
clearly resolved. The naive scan, having the same number of points
as the other two results, fails to see such detail. 
\footnote{There is only one point
of large cross section in this area which gives a faint hint of 
what might be going on in that region of parameter space.
One would need a tremendous amount of additional sampled points in order to
resolve it with the naive approach.}
These areas of large cross section actually correspond to the large 
cross section area for low $\MSF$ in Figures
\ref{1dplots-log}(b) and \ref{1dplots-lin}(b).

\begin{figure}[tb]
{\setlength{\unitlength}{1cm}
\begin{picture}(15,7)(0,0)

\put(0.0,0.2){\resizebox*{.43\width}{.43\height}{\includegraphics*{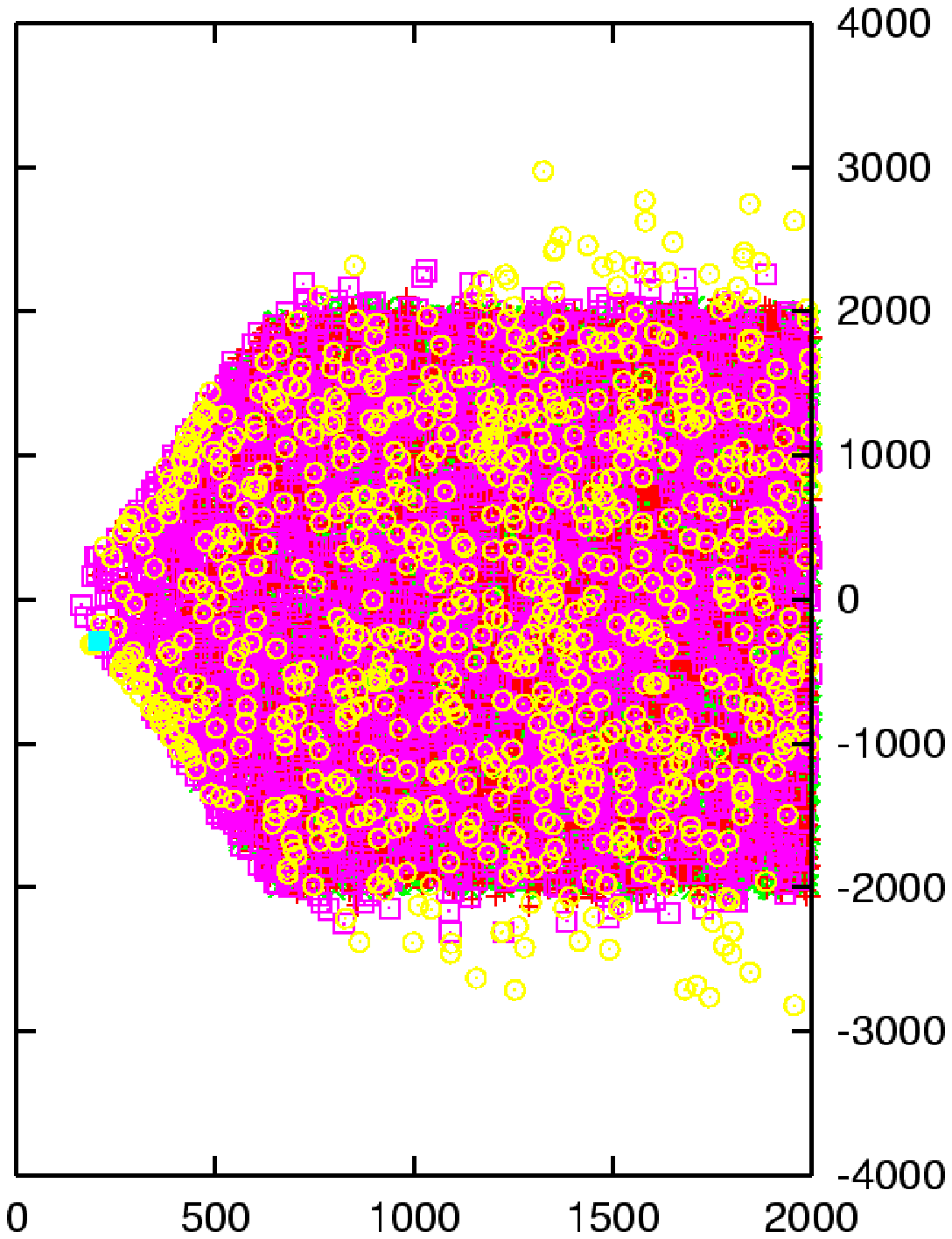}}}
\put(.4,6.1){\tiny naive}
\put(5.0,0.2){\resizebox*{.43\width}{.43\height}{\includegraphics*{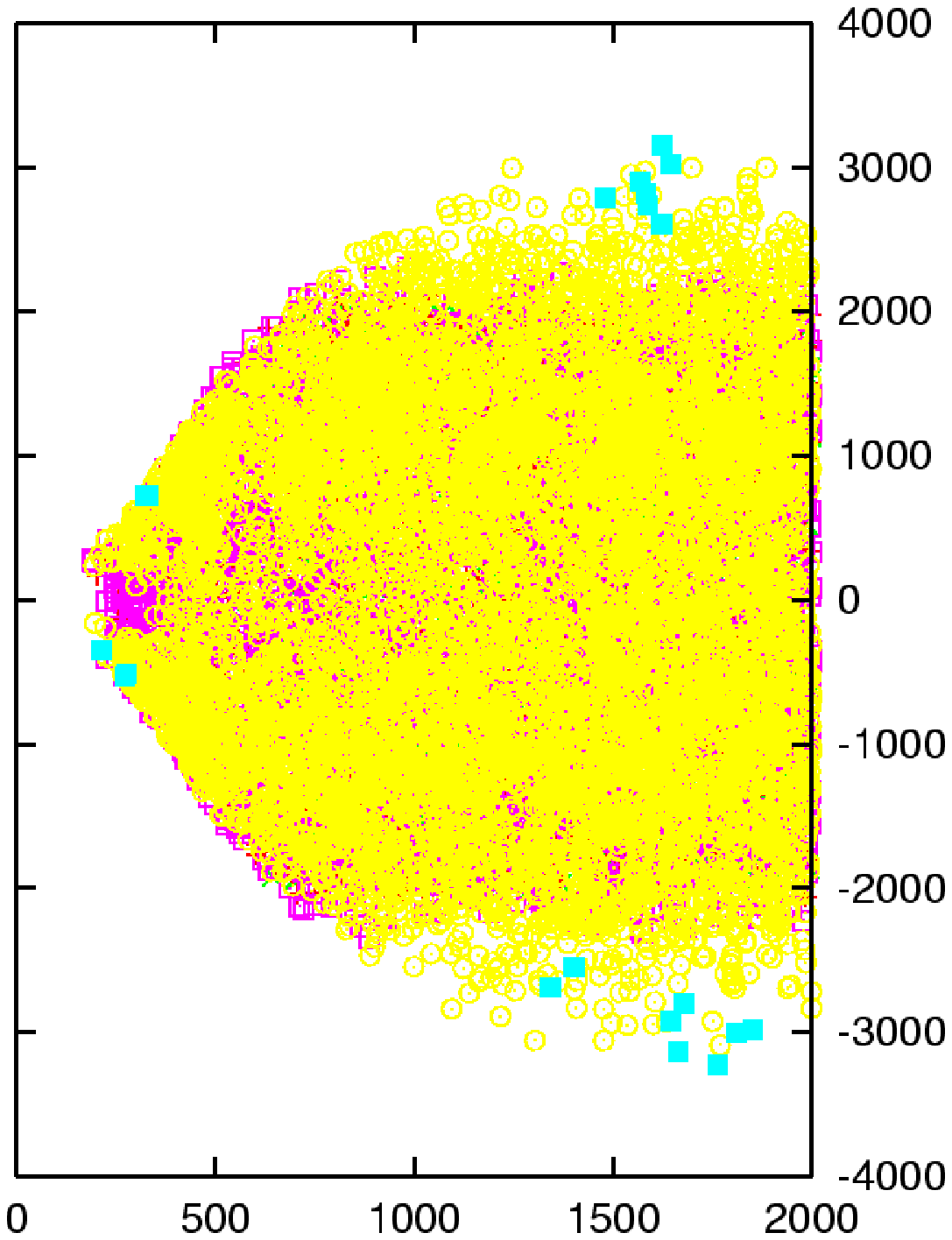}}}
\put(5.4,6.1){\tiny adaptive, $G=\sigma$}
\put(10.0,0.2){\resizebox*{.43\width}{.43\height}{\includegraphics*{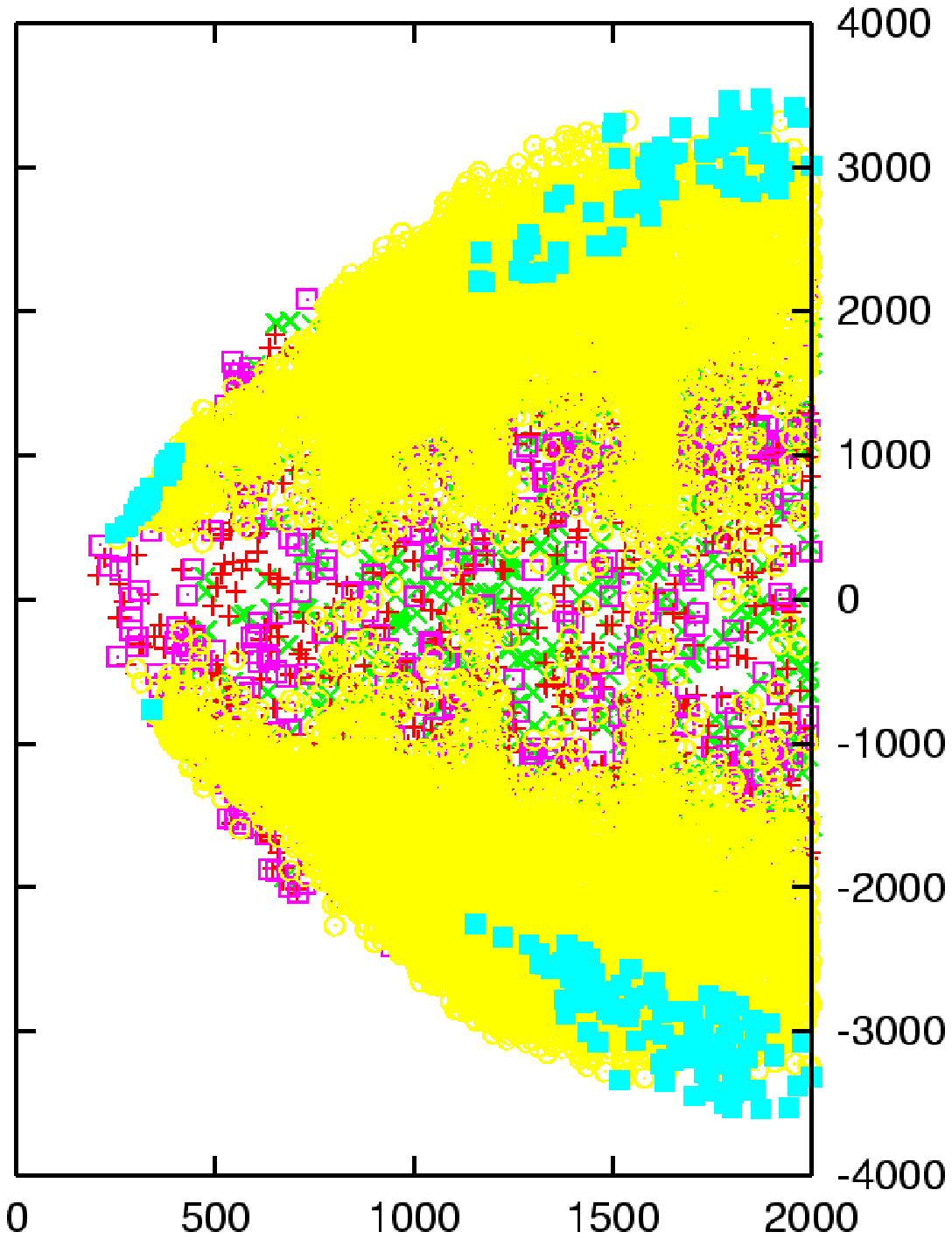}}}
\put(10.4,6.1){\tiny adaptive, $G=\theta(\sigma-\sigma_0)\,\sigma$}

\put(.4,1.9){\tiny $\sigma\; [\pb]$}
\put(5.4,1.9){\tiny $\sigma\; [\pb]$}
\put(10.4,1.9){\tiny $\sigma\; [\pb]$}

\put(.4,1.65){\tiny\GNUPlotB{$0 - \E{-7}$}}
\put(.4,1.4){\tiny\GNUPlotA{$\E{-7} - \E{-6}$}}
\put(.4,1.2){\tiny\GNUPlotD{$\E{-6} - \E{-5}$}}
\put(.4,1.0){\tiny\GNUPlotF{$\E{-5} - \E{-4}$}}
\put(.4,0.8){\tiny\GNUPlotE{$\E{-4} - \E{-3}$}}

\put(5.4,1.65){\tiny\GNUPlotB{$0 - \E{-7}$}}
\put(5.4,1.4){\tiny\GNUPlotA{$\E{-7} - \E{-6}$}}
\put(5.4,1.2){\tiny\GNUPlotD{$\E{-6} - \E{-5}$}}
\put(5.4,1.0){\tiny\GNUPlotF{$\E{-5} - \E{-4}$}}
\put(5.4,0.8){\tiny\GNUPlotE{$\E{-4} - \E{-3}$}}

\put(10.4,1.65){\tiny\GNUPlotB{$0 - \E{-7}$}}
\put(10.4,1.4){\tiny\GNUPlotA{$\E{-7} - \E{-6}$}}
\put(10.4,1.2){\tiny\GNUPlotD{$\E{-6} - \E{-5}$}}
\put(10.4,1.0){\tiny\GNUPlotF{$\E{-5} - \E{-4}$}}
\put(10.4,0.8){\tiny\GNUPlotE{$\E{-4} - \E{-3}$}}

\put(1.8,0){\small $\MSF\;[\gev]$}
\put(6.8,0){\small $\MSF\;[\gev]$}
\put(11.8,0){\small $\MSF\;[\gev]$}
\put(14.5,3.45){\small $X_t\;[\gev]$}
\end{picture}
}
\caption{\label{2dplot}
Cross section values plotted versus $\MSF$ and $X_t$
in energy bins obtained by
the three methods of scanning the parameter space.}
\end{figure}

Figure \ref{2dplot} shows the cross section versus the two important 
parameters $\MSF$ and $X_t$. 
As all three scans have roughly the same number of sampled points,
the improvement in finding parameter regions of large cross section
for the adaptive scans is obvious.
But, it should be mentioned that the results presented here 
are meant to demonstrate the performance of the different methods.
For a physics study the amount of sampled points would be much higher
and also a combination of scans with different importance functions is
conceivable.

\begin{figure}[tb]
{\setlength{\unitlength}{1cm}
\begin{picture}(10,6.5)(0,0)
  \psfrag{naive}[l][l]{\LARGE naive}
  \psfrag{ADGISSIGMA}[l][l]{\LARGE adaptive, $G=\sigma$}
  \psfrag{ADGISTHETASIGMA}[l][l]{\LARGE adaptive, $G=\theta(\sigma-\sigma_0) \sigma$}
  \psfrag{M_sfermion}[c][b]{\LARGE $\MSF [\gev]$} 
  \psfrag{ADAPTIVERADIKAL}{\Large adaptive, $G=\theta(\sigma-\sigma_0)\,\sigma$}
  \psfrag{ADAPTIVE}{\Large adaptive, $G=\sigma$}
  \psfrag{NAIVE}{\Large naive}
\put(2,0){\resizebox*{.6\width}{.6\height}{\includegraphics*{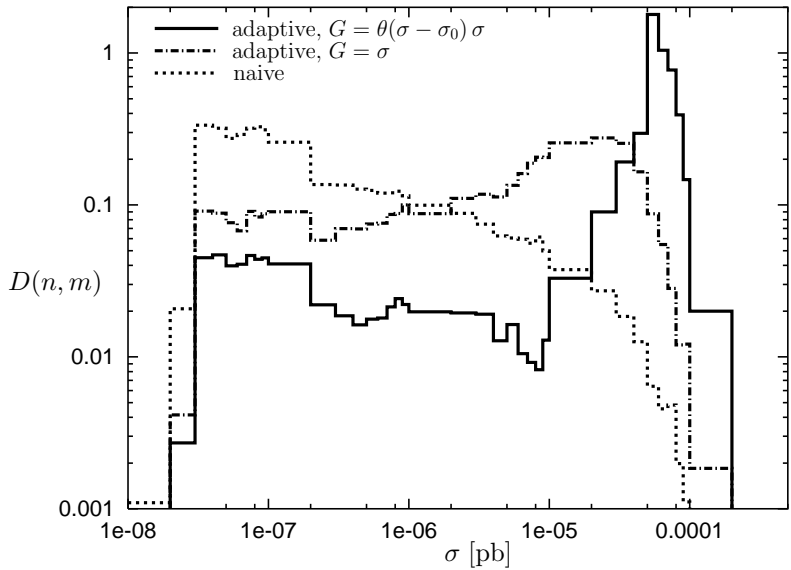}}}
\put(1.8,3.5){\small $D(n,m)$}
\put(7.6,-.1){\small $\sigma\;[\pb]$}
\end{picture}
}
\caption{\label{pointdensities}
Relative densities of sampled points in the bins
$[m\TE{-n},(m+1)\TE{-n}]\;\text{picobarn}$, with  $n\in\{3,\ldots,8\},
m\in\{1,\ldots,9\}$ for the three parameter scan approaches.}
\end{figure}

A quantitative view on the performance of the different approaches is 
given in Figure \ref{pointdensities}, which shows for each of 
the three approaches the relative density $D(n,m)$
of sampled points in cross section bins 
$[m\TE{-n},(m+1)\TE{-n}]\;\text{picobarn}$, with  $n\in\{3,\ldots,8\},
m\in\{1,\ldots,9\}$.  
$D(n,m)$ is defined as the number of points per bin, $N(n,m)$, normalized to
the total number of sampled points, 
\begin{align*}
D(n,m) &= \frac{N(n,m)}{\sum_{n,m} N(n,m)} \;.
\end{align*}
In the naive approach the density of sampled points is just
the distribution of cross section values which emerges for a random
choice of parameter values in the given hypercube of scan parameters.
The adaptive approaches gain 
relative densities of sampled points 
in the large cross section area which are 
one to two orders of magnitude higher than for the naive approach.

The method of adaptive scanning has been successfully applied in two 
other analyses \cite{nuNucleon,jaeger-rosiek}.
The authors of \cite{jaeger-rosiek} performed a parameter scan over a 
16--dimensional subset of the MSSM parameter space.
Encouraged by the results, they even performed a more general,
66--dimensional scan including many more parameters that
only indirectly influence their observables, obtaining
results consistent with the smaller scan.

\section{Summary}\label{summary}
We present a new method to perform parameter scans for theoretical
predictions: ``adaptive scanning''.
The method is designed to handle parameter spaces of large dimensionality
and is very efficient if there is an emphasis on certain features of the
prediction. 
Furthermore, in cases where parts of the 
parameter space are excluded or forbidden by further restrictions our
method of scanning can focus adaptively on the allowed region.
The scan of a prediction $f(p_i)$ over (a portion of) a parameter space $\{p_i\}$
with emphasis on certain features is realized by
defining 
an importance 
function $G(f,p_i)$ and 
sampling adaptively more points
in regions where $G$ is large. 
This is done by numerically estimating the integral of $G$ over 
(a portion of) parameter space with an adaptive Monte Carlo integration
algorithm which is based on importance sampling
and by recording the sampled points.
The method can be straightforwardly implemented as a computer program using
VEGAS with a suitable random number generator 
and a function to be called by 
VEGAS 
which calculates the desired 
function $G$
and stores the sampled parameter points. 
The method 
is modular and flexible:
One is free to choose the function $G$ and it is simple to 
use different integration algorithms.
Any sophistication present in implementations of adaptive integration algorithms,
for instance parallelization, may be exploited for our method
of parameter scanning.
An exemplary FORTRAN code using VEGAS
is available from
the author upon request.

\bigskip

\section*{Acknowledgment}
I thank the authors of \cite{jaeger-rosiek}, especially Sebastian J\"ager, 
for taking over the proposal described here for  
their own analysis.
Furthermore, I thank W.~Bernreuther and S.~J\"ager for discussions and reading the
manuscript.

\end{document}